\documentclass[12pt, a4paper]{article}

\usepackage[utf8]{inputenc}
\usepackage{amsmath, amssymb, amsfonts}
\usepackage{graphicx}
\usepackage{booktabs}
\usepackage{geometry}
\usepackage{hyperref}
\usepackage{caption}

\hypersetup{
    colorlinks=true,
    linkcolor=blue,
    citecolor=blue,
    urlcolor=blue
}

\title{\Large\sc  {Consistent control of energy dissipation in non-spherical particle contact via a structure-preserving formulation}}
\author{
    \textsc{Y.T. Feng\footnote{e-mail: y.feng@swansea.ac.uk}} \\
    \small Zienkiewicz Institute for Modelling, Data and AI \\\small Faculty of Science and Engineering\\
    \small Swansea University, Swansea SA1 8EN, United Kingdom
}
\date{}

\begin{document}

\maketitle

\begin{abstract}
The control of energy dissipation in non-spherical particle contact remains an unresolved problem. Unlike spherical contact, where the interaction reduces to a one-dimensional normal oscillator, both the effective inertia and the effective stiffness depend on the evolving contact geometry, and the impact dynamics are intrinsically coupled across translational, rotational, and tangential directions. Classical damping formulations are therefore structurally incompatible with the contact dynamics they are intended to represent.

This work addresses the problem from first principles. By projecting the dynamics onto contact degrees of freedom, the interaction is shown to be governed by an instantaneous contact dynamics with a configuration-dependent projected mass and intrinsic translational--rotational coupling. Building on the exact energy--phase transformation for monotone conservative contact, we show that consistent dissipation requires a unique damping structure aligned with the underlying contact energy.

The analysis leads to two central consequences. First, the admissible damping law is not empirical but fixed by the harmonic structure revealed in transformed space. Second, the appropriate coefficient of restitution for non-spherical particles is the contact-point restitution $e_{cn}$, whereas the total energy restitution $e_E$ is a geometry-dependent outcome that includes coupling-induced energy transfer.

Numerical evidence based on smooth single-contact impacts confirms the theory: the resulting formulation controls $e_{cn}$ consistently across impact configurations, while the apparent variability of $e_E$ follows directly from the coupled dynamics. 


\vspace{0.5cm}
\end{abstract}

\newpage

\section{Introduction}

The control of energy dissipation in non-spherical particle contact remains a fundamental unresolved problem in the discrete element method (DEM). For spherical particles, the normal interaction reduces to a one-dimensional oscillator, so dissipation can be calibrated through a constant effective mass and stiffness. For non-spherical particles, however, both quantities depend on the evolving contact geometry, giving rise to intrinsically coupled translational, rotational, and tangential dynamics. Classical damping formulations are therefore unable to deliver consistent energy loss or prescribed restitution in any geometry-independent sense.

In practice, dissipation is commonly characterised through the coefficient of restitution $e$, a scalar measure of energy loss in impact. Since the introduction of DEM by Cundall and Strack \cite{Cundall1979}, restitution has typically been calibrated using the linear spring--dashpot (LSD) model, which yields the exact relationship
\begin{equation}
	e = \exp\left(-\frac{\pi \xi}{\sqrt{1-\xi^2}}\right)
\end{equation}
between $e$ and the damping ratio $\xi$. The relationship between kinematic restitution and continuous contact damping was first formalised by Hunt and Crossley \cite{Hunt1975} and later extended to general multibody systems by Lankarani and Nikravesh \cite{Lankarani1990}. 

Within the DEM community, the mathematical consistency of these normal force and damping models has been extensively reviewed (e.g., by Kruggel-Emden et al. \cite{KruggelEmden2007} and Feng \cite{Feng2023}). For spherical particles, this calibration is exact because the interaction reduces to a linear oscillator with constant effective mass and constant stiffness \cite{Cundall1979, Tsuji1992}. This has allowed extensive experimental characterisation of the baseline restitution for granular spheres \cite{Labous1997, Lorenz1997, Foerster1994}, alongside theoretical refinements for viscoelastic spherical contact \cite{Schwager2008, Kuwabara1987, Brilliantov1996}. 

Significant progress has been made on specific aspects of this problem. Tsuji, Tanaka and Ishida \cite{Tsuji1992} introduced a penetration-dependent damping model for Hertzian contact, and Antypov and Elliott \cite{Antypov2011} provided an exact calibration for the Hertz case ($p = 3/2$). More recently, the present author \cite{Feng2026a} derived a universal calibration relationship for the entire power-law family (with $p$ as the exponent),
\begin{equation}
	\label{eq:alpha_q}
\alpha = \sqrt{2(p+1)}\cdot\frac{-\ln e}{\sqrt{\pi^2+\ln^2 e}},
\end{equation}
and showed \cite{Feng2026b} that this result follows from an exact energy--phase transformation that maps any monotone contact potential onto a harmonic oscillator. These developments establish a consistent framework for dissipation control in systems governed by known one-dimensional force laws.

However, a more fundamental difficulty has been overlooked. The analysis above assumes that the contact dynamics are one-dimensional, governed by a single degree of freedom with a scalar reduced mass $m_{red}$ and a scalar stiffness $k_{\text{eff}}$. For non-spherical particles, and more generally for non-central rigid-body impacts, the contact-centric dynamics are inherently coupled: translation and rotation interact through the lever arms at the contact point, and energy is exchanged between these modes during the collision \cite{Gilardi2002, Stronge2000, Brach1991}. This coupling leads to two consequences that are not captured by existing treatments:

(i) The effective mass tensor projected onto the contact normal --- $m_n^*$ --- is not constant. It evolves throughout the collision as the contact point migrates, giving rise to a \textit{breathing mass} that is not accounted for in classical LSD formulations.

(ii) Even if the damping is calibrated to control the normal contact-point velocity, the post-collision kinetic energy includes a rotational component that is not directly affected by the normal dissipation. The commonly used definition
\begin{equation}
e = \sqrt{\frac{K_{E,f}}{K_{E,0}}}
\end{equation}
therefore measures a combination of material dissipation and geometric energy transfer, conflating two distinct physical mechanisms.

The present work addresses this problem from first principles. We show that a consistent description of dissipation requires (i) projection of the dynamics onto contact degrees of freedom, (ii) preservation of the underlying energy structure, and (iii) a redefinition of restitution in terms of contact-point kinematics. The resulting framework reveals that restitution is not a scalar material parameter for non-spherical particles, but a contact-coordinate quantity, whose apparent variability arises from translational--rotational coupling.


This paper therefore follows the structural analysis to its conclusion. Starting from the failure of classical damping (\S2), we develop a contact-centric formulation that exposes the underlying coupling structure (\S3), introduce a structure-preserving damping law consistent with this coupling (\S4), and arrive at the central result that the coefficient of restitution for non-spherical particles must be defined at the contact point (\S5). Numerical results (\S6) are restricted to smooth, analytically tractable single-contact impacts whose role is to verify the mechanics: they confirm that the resulting contact-point restitution $e_{cn}$ is controlled to within $1\%$ across a wide range of geometries and impact angles in the physically relevant stiffness regime, while the total energy restitution $e_E$ departs through a coupling-induced energy transfer that is captured exactly by an impulsive closed-form solution.

\section{The failure of classical damping}

In discrete element simulations, normal contact dissipation is typically modelled using a viscous dashpot with coefficient
\begin{equation}
	\eta = 2\xi\sqrt{m_{\text{red}}\,k_{\text{eff}}},
\end{equation}
where $m_{\text{red}}$, defined by
\begin{equation}
	\frac{1}{m_{\text{red}}}=\frac{1}{m_1}+\frac{1}{m_2},
\end{equation}
is the reduced mass of the two particles with masses $m_1$ and $m_2$; $k_{\text{eff}}$ is the contact stiffness; and $\xi$ is determined from the target coefficient of restitution. For the classical linear spring--dashpot model,
\begin{equation}
	F_n = k_n \delta,
\end{equation}
so that
\begin{equation*}
	k_{\text{eff}} = k_n = \text{const}.
\end{equation*}
For spherical particles, this yields an exactly linear oscillator with constant reduced mass and constant stiffness.

For Hertzian contact, by contrast, the stiffness is nonlinear and the exact mapping between damping and restitution requires the correction derived by Antypov and Elliott \cite{Antypov2011}. Parallel developments in the physics community (e.g., Brilliantov et al. \cite{Brilliantov1996}) have derived damping models for viscoelastic spheres, though these intentionally yield velocity-dependent restitution rather than a prescribed kinematic parameter.

For non-spherical particles --- increasingly modelled via superquadrics \cite{Cleary2010, Podlozhnyuk2017}, multi-sphere clumps, or polyhedra \cite{Lu2015} --- the classical assumption of a fixed linear oscillator breaks down. This is particularly evident within the Energy-Conserving Contact (ECC) framework \cite{Feng2021a, Feng2021b, Feng2021c}, developed for arbitrarily shaped particles, where the contact interaction is derived from a scalar energy potential rather than prescribed empirically. The interaction is governed by an evolving local geometry, and the corresponding effective inertia and stiffness are therefore no longer constants.

More fundamentally, the reduced mass $m_{\text{red}}$ is not, in general, the correct inertial quantity, because the normal force typically acts at a finite lever arm from the mass centres and therefore couples translation and rotation. The correct object is the effective mass tensor at the contact point, from which the projected normal mass is obtained, as developed in \S3. This projected normal mass evolves with the instantaneous contact geometry.

The difficulty is therefore structural. For non-spherical contact, the local oscillator governing dissipation is configuration-dependent: both its inertia and its stiffness vary throughout the collision. Any consistent treatment of dissipation must therefore begin with a reformulation of the dynamics at the contact level.

\section{Contact-centric dynamics}

\subsection{The contact-centric equation}

In classical formulations of particle dynamics, the equations of motion are expressed in terms of the translational and rotational degrees of freedom of each body. However, to analyse the failure identified in \S2, it is more natural to describe the interaction directly at the contact point. This requires projecting the Newton--Euler equations onto the motion associated with the contact, leading to a contact-centric representation of the dynamics. While projecting rigid-body equations onto contact constraints is an established technique in multibody dynamics and non-smooth mechanics \cite{Featherstone2014, Pfeiffer1996, Brogliato1999}, its implications for energy dissipation have not been systematically exploited in DEM.

The fundamental kinematic variable is the relative velocity at the contact point $\mathbf{x}_c$, defined as
\begin{equation}
\mathbf{v}_\text{rel} = (\mathbf{v}_2 + \boldsymbol{\omega}_2 \times \mathbf{r}_2) - (\mathbf{v}_1 + \boldsymbol{\omega}_1 \times \mathbf{r}_1),
\end{equation}
where $\mathbf{v}_i$ and $\boldsymbol{\omega}_i$ are the translational and rotational velocities of body $i$ at the mass centre respectively, and 
\begin{equation}
\mathbf{r}_i = \mathbf{x}_c - \mathbf{x}_i
\end{equation}
denotes the lever arm from the centre of mass of body $i$ to the contact point.

Differentiating with respect to time and substituting the Newton--Euler equations yields the contact-level acceleration in the form
\begin{equation}
\dot{\mathbf{v}}_\text{rel} = -\mathbf{M}_{\text{eff}}^{-1}\mathbf{F}_c + \mathbf{Q}_{\text{vel}},
\end{equation}
where $\mathbf{F}_c$ is the total contact force, $\mathbf{Q}_{\text{vel}}$ collects velocity-dependent terms arising from the rotation of the bodies, and $\mathbf{M}_{\text{eff}}$ is the effective mass tensor that maps forces at the contact point to accelerations.

The inverse effective mass tensor is given explicitly by
\begin{equation}
	\label{eq:M_eff}
\mathbf{M}_{\text{eff}}^{-1}
=
\left(\frac{1}{m_1} + \frac{1}{m_2}\right)\mathbf{E}_3
-
[\mathbf{r}_1]_{\times}\,\mathbf{I}_1^{-1}\,[\mathbf{r}_1]_{\times}
-
[\mathbf{r}_2]_{\times}\,\mathbf{I}_2^{-1}\,[\mathbf{r}_2]_{\times},
\end{equation}
where $\mathbf{E}_3$ is the identity matrix of order 3; $[\mathbf{r}]_{\times}$ denotes the skew-symmetric matrix associated with the cross product and $\mathbf{I}_i$ is the inertia tensor of body $i$ about its centre of mass.

This shows that, even for a single contact, the interaction is governed by a full tensorial mapping between force and acceleration at the contact point. As the contact point $\mathbf{x}_c$ evolves during the collision, the lever arms $\mathbf{r}_i$ change, and hence the effective mass tensor $\mathbf{M}_{\text{eff}}$ becomes explicitly configuration-dependent. This time dependence forms the basis of the dissipation inconsistency in classical formulations identified in \S2 and will be further discussed in \S3.3.

\subsection{The coupling structure}

For spheres, $\mathbf{r}_i \parallel \mathbf{n}$ (the unit contact normal) and $\mathbf{I}_i = I_i \mathbf{E}_3$, so $\mathbf{M}_{\text{eff}}^{-1}$ becomes diagonal in the normal--tangential decomposition. As a result, normal and tangential motions decouple, and $m_n^* = m_{\text{red}}$ remains constant. This decoupling is a geometric consequence of spherical symmetry: the line of action of the normal contact force passes through both mass centres, so no torque is generated and rotation is not excited.

For non-spherical particles, the off-diagonal entries of $\mathbf{M}_{\text{eff}}^{-1}$ are generally non-zero. A purely normal force induces tangential acceleration and rotation; conversely, rotational motion modifies the normal dynamics. To characterise the strength of this interaction, we introduce the dimensionless coupling index
\begin{equation}
	\kappa
	=
	\frac{|(\mathbf{M}_{\text{eff}}^{-1})_{nt}|}
	{\sqrt{(\mathbf{M}_{\text{eff}}^{-1})_{tt}(\mathbf{M}_{\text{eff}}^{-1})_{nn}}},
\end{equation}
where the subscripts $n$ and $t$ denote the projections of the inverse effective mass tensor onto the contact normal and the relevant tangential direction, respectively. This provides a convenient, coordinate-free measure of the degree of coupling. By construction, $\kappa = 0$ for spheres and for non-spherical shapes, it increases with both aspect ratio and impact obliquity in general.

\subsection{The breathing mass}

The effective mass governing the normal component of the contact interaction is given by 
\begin{equation}
	\label{eq:mn*}
	m_n^* = \left(\mathbf{n}^T \mathbf{M}_{\text{eff}}^{-1} \mathbf{n}\right)^{-1}.
\end{equation}
It is evident from eq.~(\ref{eq:M_eff}) that unlike the reduced mass used in classical formulations, $m_n^*$ depends on the contact configuration.

The physical origin of this configuration dependence is the offset between the line of action of the normal contact force and the mass centres of the contacting bodies. When the normal force does not pass through the mass centre of either body, it generates a torque that couples translation and rotation. The projected normal mass $m_n^*$ (eq.~(\ref{eq:mn*})) then becomes a \textit{dynamic geometric state variable} --- it depends on the instantaneous lever arms, contact normal, and inertia tensors, all of which evolve as the particles penetrate and rotate. We refer to this phenomenon as the \textit{breathing mass} of the contact.

For smooth convex bodies, the lever arm $r_x$ (the component of $\mathbf{r}$ perpendicular to the contact normal) is typically near its maximum at first contact, where the contact occurs at the most offset point on the surface. As penetration deepens and the bodies reorient, the contact point moves inward, $r_x$ decreases, and $m_n^*$ increases toward the translational mass. For these cases, the breathing mass therefore starts at its minimum and increases during the collision.

This behaviour can be illustrated using an ellipsoid with semi-axes $(a,b,c)$ at orientation $\theta$ impacting a wall. At first contact, the lever arm is given by
\begin{equation}
r_x = \frac{Q_{13}\rho}{Q_{11}},
\end{equation}
where $Q_{3\times 3} =\{Q_{ij}\}$ is the rotated shape matrix and $\rho$ is the effective radius obtained from the Schur complement (see Appendix). The corresponding initial projected mass, where $I_{yy}$ is the moment of inertia about the y-axis, is
\begin{equation}
	\label{eq:mn_t0}
m_n^*(t_0) = \left(\frac{1}{m} + \frac{r_x^2}{I_{yy}}\right)^{-1},
\end{equation}
which provides a lower bound for $m_n^*$ during the contact. For an ellipsoid with $a/b = 1.67$ at $\theta = 30^\circ$, this yields
$
m_n^* = 0.648\,m,
$
approximately $35\%$ below the reduced mass used in classical DEM.

\subsection{The maximum contact frequency}

The coupling identified above has a direct dynamical consequence: it modifies the natural frequencies of the contact system. For the frictionless case (no tangential stiffness), the stiffness tensor has rank one, $\mathbf{K} = k_n\,\mathbf{n}\mathbf{n}^T$, and the system possesses a single nonzero eigenfrequency,
\begin{equation}
\omega_{\max}^2 = \frac{k_{\text{eff}}}{m_n^*}.
\end{equation}
Because $m_n^* < m_{\text{red}}$ for non-spherical particles in oblique contact, this frequency exceeds the classical estimate $k_{\text{eff}}/m_{\text{red}}$, even without tangential coupling. The breathing mass alone increases the maximum frequency.

When tangential stiffness is present (frictional contact), the stiffness tensor becomes full-rank and the off-diagonal coupling in $\mathbf{M}_{\text{eff}}^{-1}$ leads to eigenvalue repulsion, further increasing $\omega_{\max}$ beyond both $k_n/m_n^*$ and $k_t/m_t^*$ (where $k_t$ and $m_t^*$ are the stiffness and projected mass in the tangent direction respectively). A timestep chosen from decoupled estimates will be systematically too large, providing a previously unrecognised source of instability in non-spherical DEM simulations. The quantitative treatment of this effect is deferred to a companion work to be reported elsewhere.

\section{Structure-preserving dissipation}

The contact-centric formulation developed in \S3 shows that the dynamics of non-spherical contact are governed by a configuration-dependent effective mass and intrinsic coupling between degrees of freedom. In such a system, dissipation cannot be introduced as an independent empirical term without violating the underlying energy structure.

Classical damping formulations, which prescribe a viscous force proportional to velocity using constant effective mass and stiffness, are therefore inconsistent with the contact dynamics of non-spherical particles. As shown in \S\S2--3, both the stiffness and the effective mass evolve during the collision, and the interaction cannot be reduced to a fixed linear oscillator.

A consistent description of dissipation should therefore be constructed in a manner that preserves the structure of the underlying energy. Building on the energy-based coordinate transformation developed in \cite{Feng2026a, Feng2026b}, we adopt a representation in which the nonlinear contact system is mapped onto an equivalent linear oscillator, allowing dissipation to be introduced in a form that remains compatible with the evolving dynamics.

\subsection{The energy-phase transformation}

A suitable framework is provided by the energy-based transformation developed in \cite{Feng2026a, Feng2026b}, in which the nonlinear contact dynamics are mapped onto an equivalent linear oscillator through a combined coordinate transformation and time reparametrisation.

Starting from the \textit{conservative normal contact dynamics}, the interaction between two bodies is governed by a scalar potential $W(\delta)$, such that the normal contact force derives from
\begin{equation}
F_n(\delta) = W'(\delta).
\end{equation}
The equation of motion along the normal direction therefore reads
\begin{equation}
m_n^*(\delta)\,\ddot{\delta} + W'(\delta) = 0,
\end{equation}
Here, $W(\delta)$ is the \textit{contact potential energy}, representing the elastic energy stored due to interpenetration (or deformation) of the contacting bodies. It is assumed to satisfy: 1) $W(\delta) \ge 0$ with $W(0)=0)$; 2) $W'(\delta) \ge 0$ for $\delta \ge 0$ (repulsive contact);
and 3) $W(\delta)$ is monotone increasing for $\delta > 0$.

This formulation ensures that the system is conservative, with total energy
\begin{equation}
E = \tfrac{1}{2} m_n^*(\delta)\dot{\delta}^2 + W(\delta),
\end{equation}
which remains constant during purely elastic contact.

To reveal the underlying structure of this nonlinear system, we introduce the energy coordinate
\begin{equation}
x = \sqrt{\frac{2W(\delta)}{K}},
\end{equation}
where $K$ is an arbitrary positive reference stiffness. To preserve the dynamical structure, time is reparametrised according to
\begin{equation}
\frac{d\tau}{dt} = \sqrt{\frac{M}{m_n^*(\delta)}}\,\frac{dx}{d\delta},
\end{equation}
where $M$ is an arbitrary positive reference mass. Under this transformation, the governing equation reduces exactly to
\begin{equation}
M\frac{d^2x}{d\tau^2} + Kx = 0,
\end{equation}
which is a linear harmonic oscillator.

This transformation is exact for any monotone contact potential $W(\delta)$, and the auxiliary constants $K$ and $M$ cancel from all physically measurable quantities. The nonlinear contact system is therefore dynamically equivalent to a linear oscillator in the transformed variables. It shows that the apparent nonlinearity of contact dynamics is not intrinsic, but arises from the choice of coordinate \cite{Feng2026b}.

Within this representation, the admissible form of dissipation is dictated by the requirement of precise control over energy loss. This requires the coefficient of restitution to be independent of impact velocity. In the transformed coordinates, this is achieved only if the damping remains linear in the oscillator variables; otherwise, amplitude dependence is introduced and the restitution becomes velocity-dependent. This condition uniquely determines the corresponding damping law in the original variables,
\begin{equation}
C(\delta) \propto \frac{W'(\delta)}{\sqrt{W(\delta)}},
\end{equation}
which is the unique form consistent with velocity-independent restitution. The corresponding calibration is
\begin{equation}
\xi =\xi_L \equiv \frac{-\ln e}{\sqrt{\pi^2+\ln^2 e}}.
\end{equation}
For the power-law family, this yields
\begin{equation}
	C \propto \delta^{(p-1)/2}
\end{equation}
and the universal calibration \cite{Feng2026a}
\begin{equation}
	\alpha = \sqrt{2(p+1)}\,\xi_L.
\end{equation}
For the Hertz case $(p=3/2)$, this reduces to the damping form
\begin{equation}
	C \propto \delta^{1/4},
\end{equation}
used by Tsuji et al.\ \cite{Tsuji1992}, and to the Hertz-specific calibration
\begin{equation}
	\alpha = \sqrt{5}\,\xi_L,
\end{equation}
as obtained by Antypov and Elliott \cite{Antypov2011}.

\subsection{Realisation for arbitrary conservative normal contact laws}

The preceding structure is general and applies to any conservative normal contact law with potential $W(\delta)$ and force $F_n = W'(\delta)$.
Writing the universal damping law in equivalent linear spring--dashpot form identifies the instantaneous effective stiffness as \cite{Feng2026b}
\begin{equation}
	k_{\text{eff}}(\delta)=\frac{[W'(\delta)]^2}{2W(\delta)}.
\end{equation}

For ECC Model I with the linear contact-volume law \cite{Feng2021b},
\begin{equation}
	W = K_n V_c, \qquad F_n = K_n S_n,
\end{equation}
where $V_c$ and $S_n$ are the contact volume and area respectively. Since the force is expressed through the evolving contact area, the corresponding stiffness must also be updated dynamically.
Along the normal contact path, with $\delta$ defined as the normal overlap coordinate, the geometric identity
\begin{equation}
	\frac{dV_c}{d\delta}=S_n,
\end{equation}
holds.
Hence
\begin{equation}
	W'(\delta)=K_n\frac{dV_c}{d\delta}=K_nS_n,
\end{equation}
and the instantaneous effective stiffness becomes
\begin{equation}
	k_{\text{eff}}(t)=K_n\,\frac{S_n(t)^2}{2V_c(t)},
\end{equation}
which reduces to $2\pi R K_n$ for a sphere.

The adaptive damping coefficient is therefore evaluated as
\begin{equation} \label{eq:adaptive_eta}
	\eta(t)=2\xi\sqrt{m_n^*(t)\,k_{\text{eff}}(t)},
\end{equation}
in which both $m_n^*$ and $k_{\text{eff}}$ are configuration-dependent quantities. In this way, the universal law is applied consistently with the evolving contact geometry.

\subsection{Scope of the instantaneous approximation}

The formulation above is exact when $m_n^*$ remains constant, as in aligned impacts or spherical contact. For non-spherical particles undergoing oblique impact, $m_n^*$ varies during the collision, and the instantaneous application of the damping law constitutes an approximation.

The accuracy of this approximation is governed by the relative penetration $\delta_{\max}/\rho$, where $\rho$ denotes a characteristic local radius of curvature at the contact. This ratio measures the extent to which the contact geometry evolves during the collision. When $\delta_{\max}/\rho$ is small, the variation of the effective mass and stiffness remains limited, and the instantaneous application of the damping law provides an accurate approximation of the exact behaviour.

For typical DEM stiffnesses, where $\delta_{\max}/\rho < 2\%$, the approximation is therefore highly accurate. A quantitative assessment is provided in \S6.5.

\section{The coefficient of restitution for coupled contact}

\subsection{The ambiguity}

For frictionless spherical particles, there is a single kinetic degree of freedom at the contact: the normal approach velocity $v_n$. The coefficient of restitution is therefore unambiguous, with equivalent definitions
\begin{equation}
e = \frac{|v_n^{\text{exit}}|}{|v_n^{\text{entry}}|} = \sqrt{\frac{K_{E,f}}{K_{E,0}}},
\end{equation}
since all kinetic energy resides in the normal mode.

For non-spherical particles, the contact-centric formulation (\S3) shows that energy is exchanged between normal and rotational modes during the collision through the off-diagonal coupling in $\mathbf{M}_{\text{eff}}^{-1}$. At separation, the particle carries kinetic energy in both translation and rotation, in proportions that depend on the impact geometry. As a result, the commonly used definitions of $e$ --- velocity-based, energy-based, and dissipation-based --- are no longer equivalent, giving rise to a fundamental ambiguity in the definition of restitution. As extensively discussed in classical impact mechanics \cite{Gilardi2002, Stronge2000, Brach1991, Goldsmith1960}, the kinematic (Newton), kinetic (Poisson), and energetic (Stronge) definitions of restitution are only strictly equivalent for central, collinear collisions.

\subsection{The contact-point definition}

The \textit{contact-point normal velocity}, which is the relative velocity of the material points at the contact interface projected onto the normal direction, is given by
\begin{equation}
	v_{cn} = \mathbf{n}\cdot \mathbf{v}_\text{rel} = \mathbf{n}\cdot \Big[ (\mathbf{v}_2 + \boldsymbol{\omega}_2 \times \mathbf{r}_2) - (\mathbf{v}_1 + \boldsymbol{\omega}_1 \times \mathbf{r}_1) \Big].
\end{equation}
The viscous damping force, with damping coefficient $\eta$, acts directly on this velocity in the normal direction
\begin{equation}
	F_{\text{damp}} = -\eta\,v_{cn},
\end{equation}
and dissipates energy at the rate
\begin{equation}
	\dot{E}_{\text{diss}} = \eta\,v_{cn}^2.
\end{equation}

A consistent definition of the \emph{controlled contact-level restitution} should therefore be formulated in terms of this quantity. We define
\begin{equation}
	e_{cn} = -\frac{v_{cn}^{\text{exit}}}{v_{cn}^{\text{entry}}},
\end{equation}
which (i) is directly controlled by the damping law; (ii) reduces to the classical definition for spherical particles; and (iii) remains well-defined as a contact-level restitution measure, irrespective of the strength of translational--rotational coupling.

This definition is contact-level rather than particle-level. In strongly coupled compliant impacts, translational and rotational energies may continue to exchange during the later stages of contact, so supplementary particle-level measures based on translational rebound may be introduced when required for diagnostics or comparison with simulations. Such measures, however, should be distinguished clearly from the controlled restitution $e_{cn}$.

\subsection{The impulsive relation and its specialisation to ellipsoid--wall impact}

To establish the relationship between the contact-point restitution $e_{cn}$ and the total energy restitution $e_E$, we consider the impulsive limit, in which the contact interaction is represented by an instantaneous normal impulse. The resulting relation can be written first in general contact coordinates and then specialised to the ellipsoid--wall impact used in the numerical examples. In the following relations, $(\cdot)^-$ and $(\cdot)^+$ denote the value of a kinematic quantity immediately before and after the impact, respectively.

For a frictionless impact, the normal impulse $J_n$ satisfies the restitution condition
\begin{equation}
v_{cn}^{+} = -e_{cn}\,v_{cn}^{-},
\end{equation}
so that
\begin{equation}
J_n = -(1+e_{cn})\,m_n^*\,v_{cn}^{-},
\end{equation}
where $m_n^*$ is the projected normal mass (eq.~(\ref{eq:mn*})).  

For a single rigid particle impacting a rigid wall, the post-impact translational and angular velocities are
\begin{equation}
\mathbf{v}^{+} = \mathbf{v}^{-} + \frac{J_n}{m}\,\mathbf{n},
\end{equation}
\begin{equation}
\boldsymbol{\omega}^{+}
=
\boldsymbol{\omega}^{-}
-
J_n\,\mathbf{I}^{-1}(\mathbf{r}\times\mathbf{n}),
\end{equation}
where $\mathbf{r}$ is the lever arm from the particle centre to the contact point.

The total energy restitution is then
\begin{equation}
e_E
=
\sqrt{
\frac{
\frac{1}{2}m\|\mathbf{v}^{+}\|^2
+
\frac{1}{2}\boldsymbol{\omega}^{+T}\mathbf{I}\boldsymbol{\omega}^{+}
}{
\frac{1}{2}m\|\mathbf{v}^{-}\|^2
+
\frac{1}{2}\boldsymbol{\omega}^{-T}\mathbf{I}\boldsymbol{\omega}^{-}
}
}.
\end{equation}
These expressions show that $e_E$ is determined not only by the dissipative parameter $e_{cn}$, but also by the redistribution of energy between translation and rotation induced by the impact geometry.

For the planar ellipsoid--wall impact considered in \S6, the motion is confined to the $x$--$z$ plane, the wall is fixed, and the only rotational degree of freedom is $\omega_y$. In this case, the general expressions reduce to
\begin{equation}
v_{cn}^{-} = v_z^{-} - \omega_y^{-} r_x,
\qquad
v_z^{+} = v_z^{-} + \frac{J_n}{m},
\qquad
\omega_y^{+} = \omega_y^{-} - \frac{J_n r_x}{I_{yy}}.
\end{equation}

The total energy restitution becomes
\begin{equation}
e_E
=
\sqrt{
\frac{
m(v_z^{+})^2 + I_{yy}(\omega_y^{+})^2
}{
m(v_z^{-})^2 + I_{yy}(\omega_y^{-})^2
}
}.
\end{equation}
This specialised form is used in the numerical results of \S6. The key point, however, is already evident from the general relations above: even when $e_{cn}$ is prescribed, the total energy restitution $e_E$ also depends on the impact geometry through the coupling between translation and rotation.

\subsection{The coupling offset}

The difference
\begin{equation}
\Delta e = e_E - e_{cn}
\end{equation}
measures the extent to which the total energy restitution exceeds the contact-point restitution. We refer to this excess as the \textit{coupling offset}. It arises because, during a non-spherical impact, part of the translational kinetic energy is redistributed into rotational motion through the coupled contact dynamics. As a result, the total post-impact energy can remain higher than would be inferred from the dissipation acting on the normal contact mode alone.

The magnitude of $\Delta e$ is governed by the geometry of the contact and the pre-impact kinematic state. In particular, it depends on the extent to which the effective mass tensor couples the normal contact motion to the other degrees of freedom, which is controlled by particle shape, contact configuration, impact orientation, and any pre-impact rotation. Different geometries therefore produce different values of $e_E$ even when the same contact-point restitution $e_{cn}$ is prescribed.

By contrast, $\Delta e$ may not depend directly on the contact stiffness, damping ratio, or material parameters once $e_{cn}$ is fixed. Those quantities determine the dissipation in the contact mode itself, whereas the coupling offset reflects the geometric redistribution of energy between modes. It is therefore not a dissipative effect in its own right, but a consequence of the coupled impact dynamics.

\subsection{Implications}

The interpretation --- that the measured $e$ is a material property of the contact --- is generally justified only for spherical particles and other uncoupled central-impact configurations. For non-spherical particles, the measured $e_E$ reflects a combination of material dissipation (controlled by $e_{cn}$) and geometric energy transfer (arising from the coupling). DEM simulations should therefore specify $e_{cn}$ as the input parameter, rather than $e_E$. The adaptive damping scheme (\ref{eq:adaptive_eta}) accepts $e_{cn}$ directly, while the resulting $e_E$ emerges as a prediction that varies with impact geometry, as experiments indicate.

This provides a mechanistic explanation for the well-known observation that the measured coefficient of restitution depends on impact angle for non-spherical grains \cite{Wolf2020, Gorham2000, Hastie2013, Wilson2023, Wang2020Exp}. The variation does not imply that the material properties change with angle; rather, it reflects the fact that the coupled dynamics transfer different fractions of the initial energy into rotation.

\section{Numerical evidence}

This section verifies three central claims of the paper: (i) exact energy conservation in the undamped case; (ii) accurate control of the contact-point restitution $e_{cn}$ in the damped case; and (iii) agreement between the compliant dynamics and the impulsive prediction for the total energy restitution $e_E$. The numerical evidence is intentionally restricted to smooth, analytically tractable single-contact impacts, while a broad treatment for more complex contact scenarios will be reported in the companion paper elsewhere.

\subsection{Setup}

An ellipsoid with semi-axes $(a,b,c) = (1.0, 0.6, 0.6)$ m and mass $m = 1$ kg impacts a rigid flat wall at $z = 0$ under gravity-free conditions. The ECC Model I formulation with linear contact energy
\begin{equation}
W = K_n V_c
\end{equation}
is used together with the adaptive damping law of eq.~\eqref{eq:adaptive_eta}. Time integration is performed using the second-order velocity Verlet scheme. The contact geometry (cross-sectional area, cap volume, contact point centroid, and projected normal mass) is computed analytically from the Schur complement of the rotated shape matrix (see Appendix), providing exact reference values.

\subsection{Energy conservation}

With no damping ($e_{cn} = 1$), the total energy
\begin{equation}
E = K_E + K_n V_c
\end{equation}
is monitored for the oblique impact case
\begin{equation}
\theta = 30^\circ,\qquad v_z = -1.5\ \text{m/s},\qquad \omega_y = 0,\qquad K_n = 10^6\ \text{Pa}.
\end{equation}

\begin{figure}[htbp]
    \centering
    \includegraphics[width=1.0\textwidth]{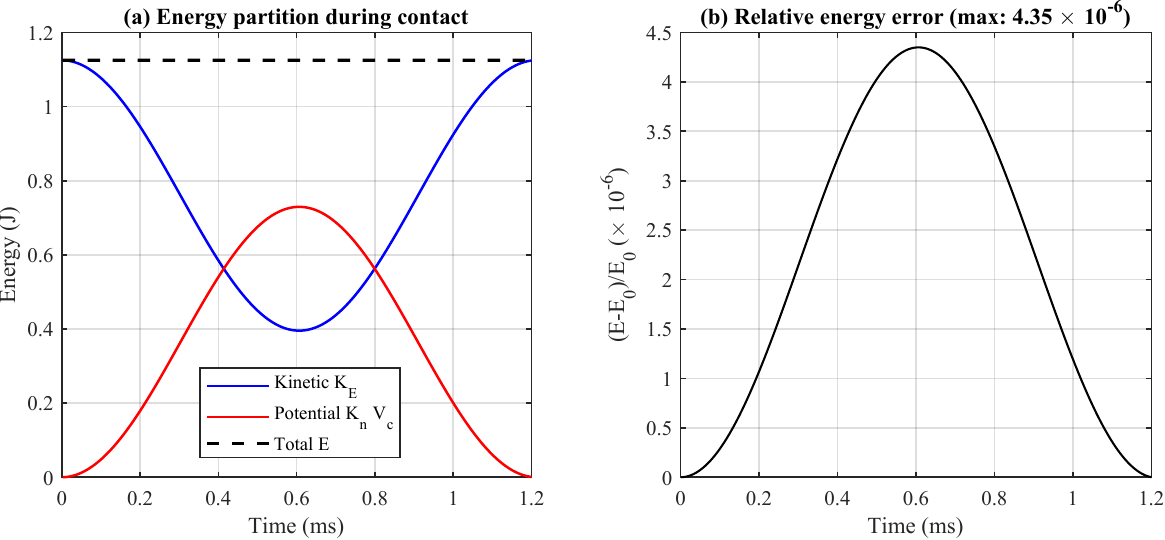}
    \caption{Energy conservation during undamped oblique impact ($\theta = 30^\circ$, $K_n = 10^6$ Pa). (a) Kinetic, potential, and total energy as functions of time. (b) Relative energy error $(E-E_0)/E_0$, showing a maximum of $4.4 \times 10^{-6}$ at peak penetration.}
    \label{fig:1}
\end{figure}

Figure~\ref{fig:1} shows the energy partition during the contact event. Panel (a) displays the exchange between kinetic and potential energy over the contact duration. Using the timestep $\Delta t = 2 \times 10^{-6}s$, panel (b) shows the relative energy error, which reaches a maximum of $4.4 \times 10^{-6}$ at maximum penetration and returns to near-zero at separation --- consistent with the symplectic character of the velocity Verlet integrator. 


\subsection{Aligned impact --- exact control}

For $\theta = 0^\circ$, the coupling vanishes and the system reduces to a one-dimensional oscillator. At $K_n = 10^3$ Pa, the achieved restitution agrees with the target to excellent accuracy (as shown in Table~\ref{tab:1}). 
In this limit, all definitions of restitution coincide. The adaptive damping law is therefore exact in the one-dimensional case, consistent with the underlying transformation theory.

\begin{table}[htbp]
\centering
\begin{tabular}{ccc}
\toprule
Target $e_{cn}$ & Achieved $e_{cn}$ & Error \\
\midrule
0.1 & 0.100 & 0.06\% \\
0.3 & 0.301 & 0.15\% \\
0.5 & 0.500 & 0.08\% \\
0.7 & 0.700 & 0.05\% \\
0.9 & 0.900 & 0.01\% \\
\bottomrule
\end{tabular}
\caption{Restitution control for aligned impact ($\theta = 0^\circ$).}
\label{tab:1}
\end{table}

\begin{figure}[htbp]
    \centering
    \includegraphics[width=1.0\textwidth]{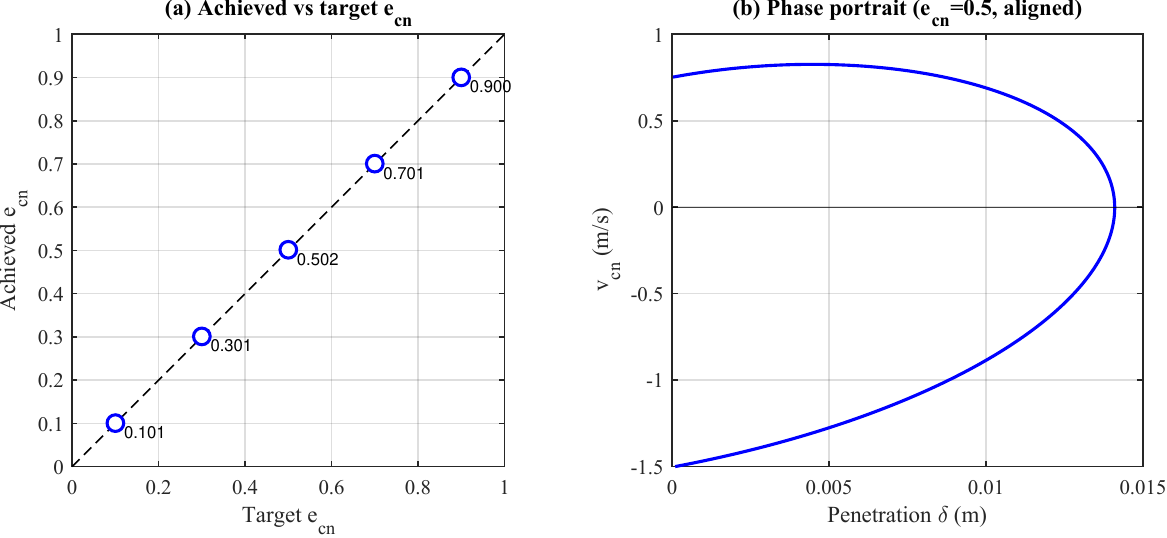}
    \caption{Aligned impact ($\theta = 0^\circ$, $K_n = 10^3$ Pa). (a) Achieved versus target $e_{cn}$. (b) Phase portrait for $e_{cn} = 0.5$, showing the damped contact oscillation.}
    \label{fig:2}
\end{figure}

Figure~\ref{fig:2}(a) confirms this by plotting achieved versus target $e_{cn}$, with all points falling on the diagonal. Figure~\ref{fig:2}(b) shows the phase portrait ($v_{cn}$ versus $\delta$) for the case $e_{cn} = 0.5$, displaying the asymmetry between approach and rebound velocities that characterises the controlled dissipation.

\subsection{Oblique impact --- $e_{cn}$ is controlled, $e_E$ departs}

For $\theta = 30^\circ$, $K_n = 10^3$ Pa, and target $e_{cn} = 0.5$, the results are summarised in Table~\ref{tab:2}. 

\begin{table}[htbp]
\centering
\begin{tabular}{lc}
\toprule
Definition & Value \\
\midrule
$e_{cn}$ (contact-point normal velocity) & \textbf{0.496} \\
$e_E$ (total energy) & 0.714 \\
Impulsive prediction for $e_E$ & 0.717 \\
\% KE in rotation at exit & 51.3\% \\
\bottomrule
\end{tabular}
\caption{Restitution definitions evaluated for oblique impact.}
\label{tab:2}
\end{table}

The adaptive scheme controls $e_{cn}$ to within $1\%$ of the prescribed value. The departure of $e_E$ from the target is not a numerical error but the coupling offset predicted by the impulsive relation. The agreement with the impulsive estimate is within $0.4\%$.

\begin{figure}[htbp]
    \centering
    \includegraphics[width=1.0\textwidth]{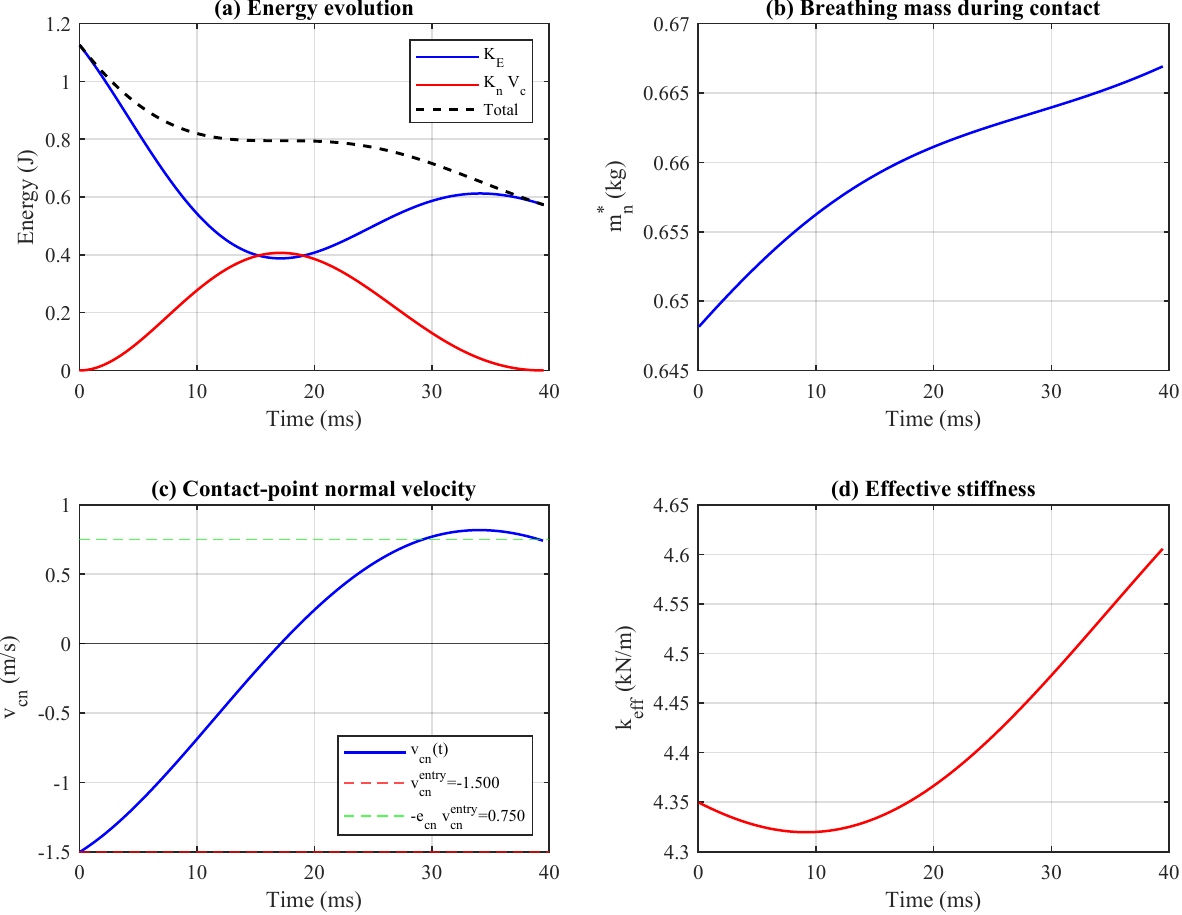}
    \caption{Oblique impact diagnostics ($\theta = 30^\circ$, $e_{cn} = 0.5$, $K_n = 10^3$ Pa). (a) Energy evolution. (b) Breathing mass $m_n^*(t)$ during contact, with the initial value (lower bound) of $0.648$. (c) Contact-point normal velocity $v_{cn}$, with entry value and target exit value marked. (d) Instantaneous effective stiffness $k_{\text{eff}}(t)$.}
    \label{fig:3}
\end{figure}

Figure~\ref{fig:3} provides detailed diagnostics of this case. Panel (a) shows the energy evolution, with the total energy decreasing by the prescribed amount while the kinetic energy at separation exceeds the target due to rotational energy. Panel (b) displays the breathing mass $m_n^*(t)$ during contact, which starts at its lower bound of $0.648\,m$ and increases as the ellipsoid rotates toward alignment. Panel (c) tracks the contact-point normal velocity, confirming that the exit value is consistent with $e_{cn} = 0.5$. Panel (d) shows the evolution of the instantaneous effective stiffness $k_{\text{eff}}$.

\subsection{The breathing mass effect}

The quality of the adaptive scheme for oblique impacts depends on how much $m_n^*$ varies during the collision. This variation is governed by the relative penetration $\delta_{\max}/\rho$, where $\rho$ is the local radius of curvature. 
A sweep over stiffness values from $10^1$ to $10^4$ Pa at $\theta = 30^\circ$ yields the results shown in Table~\ref{tab:3}.

\begin{table}[htbp]
\centering
\begin{tabular}{lcccc}
\toprule
$K_n$ (Pa) & $\delta_{\max}/\rho$ & $e_{cn}$ & Error & Rotation during contact \\
\midrule
$10^1$ & 19.7\% & 0.137 & 73\% & $33.7^\circ$ \\
$10^2$ & 6.0\% & 0.455 & 9.0\% & $8.6^\circ$ \\
$3\times10^2$ & 3.5\% & 0.484 & 3.3\% & $4.9^\circ$ \\
$10^3$ & 1.9\% & 0.495 & 1.1\% & $2.7^\circ$ \\
$3\times10^3$ & 1.1\% & 0.498 & 0.3\% & $1.6^\circ$ \\
$10^4$ & 0.6\% & 0.500 & 0.0\% & $0.85^\circ$ \\
\bottomrule
\end{tabular}
\caption{Effect of stiffness on the breathing mass approximation error.}
\label{tab:3}
\end{table}

For comparison, the aligned case at the same $K_n$ values shows errors below $0.3\%$ throughout, isolating the breathing mass as the dominant source of approximation error in the oblique case.

\begin{figure}[htbp]
    \centering
    \includegraphics[width=1.0\textwidth]{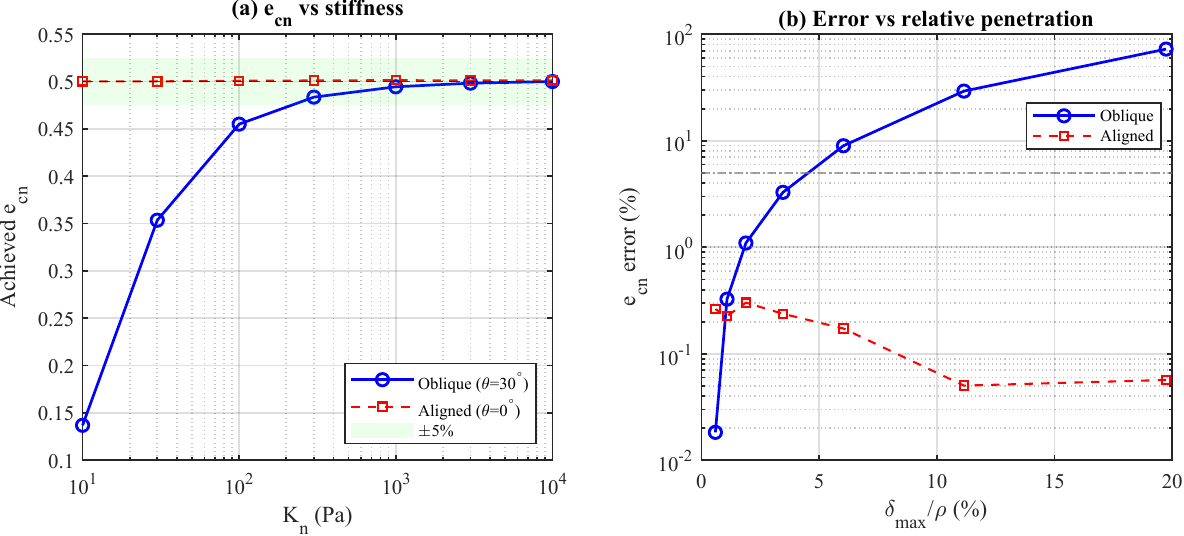}
    \caption{Breathing mass effect on restitution accuracy. (a) Achieved $e_{cn}$ versus $K_n$ for oblique ($\theta = 30^\circ$) and aligned ($\theta = 0^\circ$) impacts, with the $\pm 5\%$ band indicated. (b) Restitution error versus relative penetration $\delta_{\max}/\rho$, showing this parameter as the single governing quantity.}
    \label{fig:4}
\end{figure}

Figure~\ref{fig:4}(a) plots the achieved $e_{cn}$ against stiffness for both oblique and aligned impacts, showing the clean separation between the 1D-exact aligned case and the oblique case where the breathing mass degrades the approximation at low stiffness. Figure~\ref{fig:4}(b) replots the error against the relative penetration $\delta_{\max}/\rho$, confirming that this single dimensionless parameter governs the accuracy.

The trend is clear: for $\delta_{\max}/\rho < 2\%$, which is typical of DEM simulations with realistic stiffness, the error in $e_{cn}$ remains below $1\%$. At much softer contact levels, the error increases smoothly and remains governed by the single dimensionless parameter $\delta_{\max}/\rho$.

\subsection{Robustness across the parameter space}

To verify that accurate control of $e_{cn}$ is not specific to a single geometry, we sweep over aspect ratios $(a/b = 1.0$ to $2.5)$ and impact angles $(\theta = 0^\circ$ to $60^\circ)$ at fixed $K_n = 10^3$ Pa and target $e_{cn} = 0.5$.

The results for the \textit{aspect ratio sweep} $(\theta = 30^\circ)$ are listed in Table~\ref{tab:4}, while the results for the \textit{impact angle sweep} $(a/b = 1.67)$ are summarised in Table~\ref{tab:5}.
\begin{table}[htbp]
\centering
\begin{tabular}{ccccc}
\toprule
$a/b$ & $e_{cn}$ & $e_E$ & Coupling $\kappa$ & \% KE in rotation \\
\midrule
1.0 & 0.501 & 0.500 & 0.000 & 0.0 \\
1.4 & 0.498 & 0.625 & 0.359 & 34.7 \\
1.8 & 0.495 & 0.747 & 0.519 & 54.7 \\
2.5 & 0.505 & 0.836 & 0.611 & 53.8 \\
\bottomrule
\end{tabular}
\caption{Aspect ratio sweep at $\theta = 30^\circ$.}
\label{tab:4}
\end{table}

\begin{table}[htbp]
\centering
\begin{tabular}{cccc}
\toprule
$\theta$ & $e_{cn}$ & $e_E$ & $e_E$ (impulsive) \\
\midrule
$0^\circ$ & 0.502 & 0.500 & 0.500 \\
$15^\circ$ & 0.481 & 0.625 & 0.626 \\
$30^\circ$ & 0.495 & 0.714 & 0.717 \\
$45^\circ$ & 0.516 & 0.715 & 0.719 \\
$60^\circ$ & 0.529 & 0.658 & 0.662 \\
\bottomrule
\end{tabular}
\caption{Impact angle sweep at $a/b = 1.67$.}
\label{tab:5}
\end{table}

Two observations are central. First, $e_{cn}$ remains close to the target across all aspect ratios and impact angles, with departures of a few percent attributable to the breathing mass effect at this stiffness level ($\delta_{\max}/\rho \approx 2\%$). Second, the compliant $e_E$ agrees closely with the impulsive prediction, confirming that the coupling offset is a kinematic consequence of the impact geometry rather than a failure of the damping model.

\begin{figure}[htbp]
    \centering
    \includegraphics[width=1.0\textwidth]{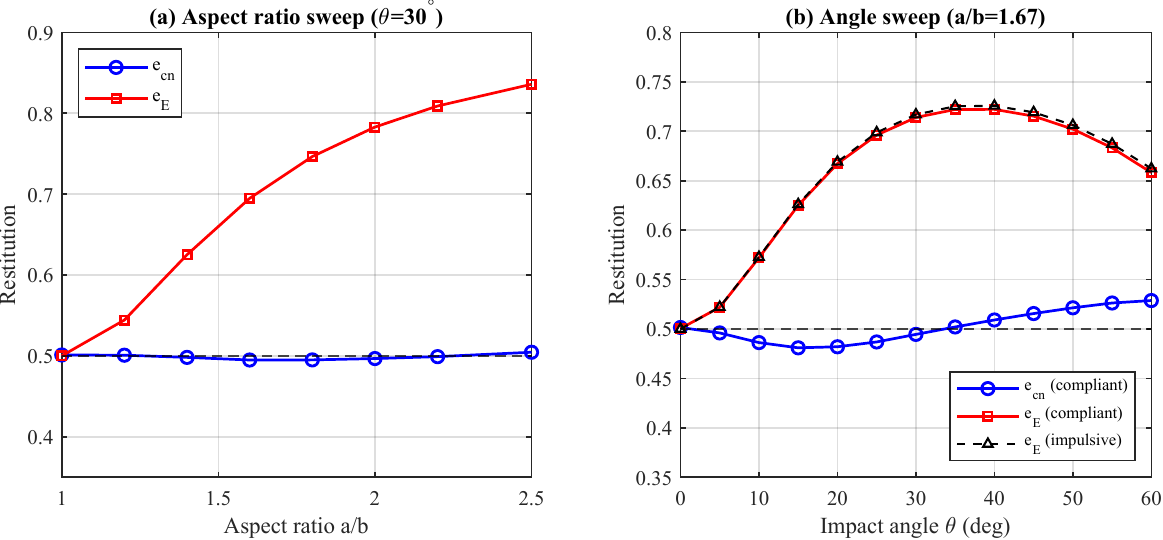}
    \caption{Robustness of $e_{cn}$ control across the parameter space ($K_n = 10^3$ Pa, target $e_{cn} = 0.5$). (a) Aspect ratio sweep at $\theta = 30^\circ$: $e_{cn}$ (blue) stays near the target while $e_E$ (red) increases with coupling. (b) Impact angle sweep at $a/b = 1.67$: $e_{cn}$ (blue), $e_E$ from compliant simulation (red), and $e_E$ from the impulsive formula (black triangles).}
    \label{fig:5}
\end{figure}

Figure~\ref{fig:5} presents these results graphically. Panel (a) shows the aspect ratio sweep: $e_{cn}$ remains near $0.5$ while $e_E$ increases steadily with aspect ratio, reaching $0.84$ at $a/b = 2.5$. Panel (b) shows the angle sweep: $e_{cn}$ stays near the target while $e_E$ traces a curve that agrees with the impulsive prediction to plotting precision.

\subsection{Convergence to the impulsive limit}

As $K_n \to \infty$, the penetration depth tends to zero, the contact duration vanishes, and the compliant solution must converge to the impulsive limit. For the undamped case ($e_{cn} = 1$), the post-impact velocities converge monotonically, as demonstrated in Table~\ref{tab:6}.

\begin{table}[htbp]
\centering
\begin{tabular}{lccc}
\toprule
$K_n$ (Pa) & $v_z^+$ error & $\omega_y^+$ error & $\delta_{\max}$ (m) \\
\midrule
$10^4$ & 2.40\% & 0.23\% & 0.0050 \\
$10^5$ & 0.75\% & 0.07\% & 0.0016 \\
$10^6$ & 0.24\% & 0.02\% & 0.0005 \\
$10^7$ & 0.08\% & 0.007\% & 0.00016 \\
\bottomrule
\end{tabular}
\caption{Convergence of compliant integration to the impulsive limit.}
\label{tab:6}
\end{table}

\begin{figure}[htbp]
    \centering
    \includegraphics[width=1.0\textwidth]{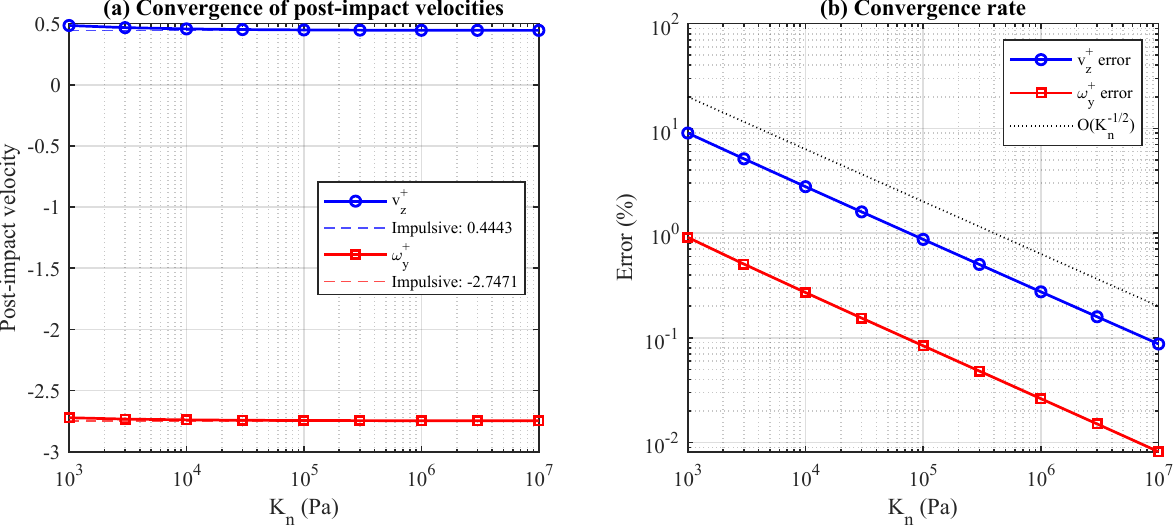}
    \caption{Convergence to the impulsive limit ($e_{cn} = 1$, $\theta = 30^\circ$). (a) Post-impact $v_z^+$ and $\omega_y^+$ versus $K_n$, with impulsive values indicated by dashed lines. (b) Relative errors on a log-log scale, showing $O(K_n^{-1/2})$ convergence.}
    \label{fig:6}
\end{figure}

Figure~\ref{fig:6}(a) shows the post-impact velocities converging toward the impulsive closed-form values (dashed lines) as $K_n$ increases. Figure~\ref{fig:6}(b) confirms the convergence rate on a log-log scale, with the errors decreasing as $O(K_n^{-1/2})$, consistent with a compliant-to-rigid convergence governed by the penetration depth.

\subsection{Full restitution sweep at different angles}

Table~\ref{tab:7} gives the total energy restitution $e_E$ achieved for different prescribed values of $e_{cn}$ at four impact angles $(a/b = 1.67,\; K_n = 10^3)$.

\begin{table}[htbp]
\centering
\begin{tabular}{ccccc}
\toprule
Target $e_{cn}$ & $\theta = 0^\circ$ & $\theta = 15^\circ$ & $\theta = 30^\circ$ & $\theta = 45^\circ$ \\
\midrule
0.1 & 0.100 & 0.444 & 0.599 & 0.602 \\
0.3 & 0.300 & 0.512 & 0.641 & 0.643 \\
0.5 & 0.500 & 0.626 & 0.717 & 0.719 \\
0.7 & 0.700 & 0.766 & 0.818 & 0.820 \\
0.9 & 0.900 & 0.920 & 0.937 & 0.937 \\
\bottomrule
\end{tabular}
\caption{Total energy restitution $e_E$ for different target $e_{cn}$ at multiple impact angles.}
\label{tab:7}
\end{table}

At $\theta = 45^\circ$, even the strongly damped case with target $e_{cn} = 0.1$ gives $e_E \approx 0.60$. This does not indicate a failure of dissipation control. Rather, it reflects the fact that a substantial fraction of the initial translational energy is transferred into rotation by the coupled dynamics, thereby limiting the total energy loss measured at the system level.

\begin{figure}[htbp]
    \centering
    \includegraphics[width=0.5\textwidth]{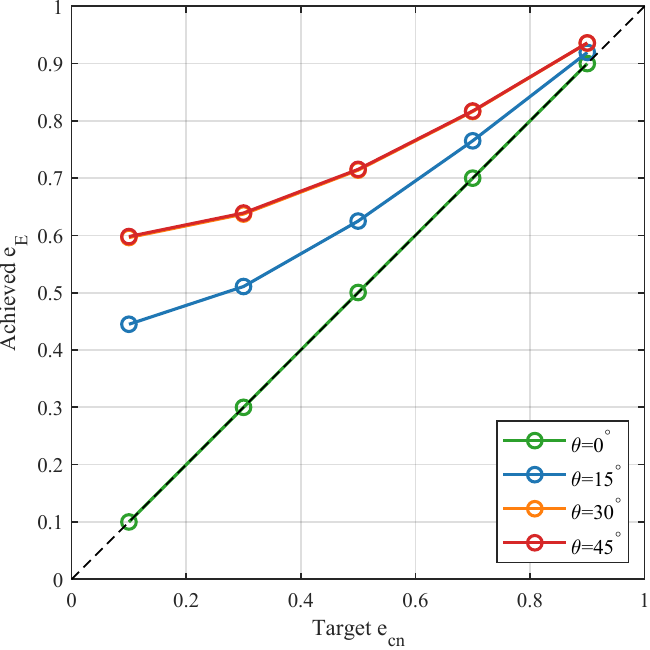}
    \caption{Total energy restitution $e_E$ versus target $e_{cn}$ at four impact angles ($a/b = 1.67$, $K_n = 10^3$ Pa). The diagonal represents perfect agreement; the vertical gap between each curve and the diagonal is the coupling offset.}
    \label{fig:7}
\end{figure}

Figure~\ref{fig:7} displays these results, plotting $e_E$ against target $e_{cn}$ for four impact angles. The $\theta = 0^\circ$ curve lies on the diagonal, confirming exact control. The oblique curves lift progressively, with the gap between each curve and the diagonal representing the coupling offset at that angle. Note that the curves for $\theta=30^\circ$ and $45^\circ$ virtually coincide (see Table~\ref{tab:7} for a precise numerical comparison).

\section{Discussion}

\subsection{Restitution is not a pure material parameter}

The central finding of this paper is that the coefficient of restitution, as conventionally defined and measured, is not a pure material property for non-spherical particles. Rather, it is a composite quantity that combines two physically distinct effects:

(a) \textit{Material dissipation}, represented by $e_{cn}$ --- the fraction of normal contact-point kinetic energy retained after the dissipative contact interaction. This is the quantity directly controlled by the damping law and provides the appropriate material measure of dissipation.

(b) \textit{Geometric energy transfer}, arising from the translational--rotational coupling encoded in the effective mass tensor $\mathbf{M}_{\text{eff}}$. This is a geometric and kinematic effect determined by the impact configuration.

The classical definition of the total energy restitution $e_E$ combines both effects. For spheres, where the coupling vanishes, $e_E = e_{cn}$. For non-spherical particles, however, coupling redistributes part of the initial translational energy into rotation. In the frictionless single-contact impacts studied here, $e_E > e_{cn}$. The apparent increase in restitution is therefore not evidence of weaker dissipation, but a consequence of the coupled impact dynamics.

\subsection{Connection to experiments}

Many particle rebound experiments infer restitution primarily from centre-of-mass translational kinematics, yielding an apparent restitution measure that is often closer to a system-level energy ratio $e_E$ than to the contact-point normal restitution $e_{cn}$, as in the experiments reported by \cite{Wolf2020, Gorham2000, Hastie2013, Dong2006, Higham2019}. The ambiguity is particularly acute when rotational degrees of freedom are not fully resolved in the measurements \cite{Higham2019}. Under such circumstances, the observed restitution often varies with impact angle and is commonly attributed to friction, roughness, or geometric effects.

The present analysis provides a precise mechanical explanation. The translational--rotational coupling embodied in $\mathbf{M}_{\text{eff}}^{-1}$ varies with the impact geometry, and therefore so does the total energy restitution $e_E$, even when the contact-point restitution $e_{cn}$ remains fixed. 
This suggests a testable hypothesis: for sufficiently smooth, single-contact impacts of a particle with known shape, much of the angular dependence of the measured restitution may be captured using a single fitted value of $e_{cn}$, with remaining scatter attributable to roughness, friction, orientation variability, and other non-ideal effects.

\subsection{The breathing mass and the scope of the approximation}

The adaptive scheme applies the universal damping law instantaneously using the current value of $m_n^*(t)$. This is exact for aligned impacts, where $m_n^*$ remains constant, and approximate for oblique impacts, where the projected mass evolves during the contact.

The quality of the approximation is governed by the relative penetration $\delta_{\max}/\rho$, which measures how much the contact geometry changes during the collision. For penetration levels typical of DEM simulations, where $\delta_{\max}/\rho < 2\%$, the resulting error in $e_{cn}$ remains below $1\%$. At extreme softness, where $\delta_{\max}/\rho \approx 5\%$ and the contact geometry evolves more substantially during impact, the error rises to about $7\%$. Even in this regime, the method remains far more reliable than fixed-coefficient damping models, which offer no comparable error control for non-spherical contact.

The impulsive value of $m_n^*(t_0)$ (eq.~(\ref{eq:mn_t0})) at first contact provides a lower bound on the breathing-mass trajectory. The relative variation $\Delta m_n^*/m_n^*$ during the collision therefore offers a direct diagnostic of the quality of the instantaneous approximation.

\subsection{Recommendations for DEM practice}

Three practical consequences follow.

First, conservative contact models should specify $e_{cn}$, not $e_E$, as the dissipation input parameter. The adaptive damping scheme acts directly on the contact-point normal mode and therefore controls $e_{cn}$, not the total energy restitution.

Second, the total energy restitution $e_E$ should be regarded as a simulation output rather than an input. Its variation with impact geometry is a physical consequence of the coupled contact dynamics, not a numerical artefact.

Third, validation against experiments should be performed at the level of the predicted function $e_E(\theta)$, rather than by attempting to assign a single geometry-independent restitution value. The theory predicts that a single fitted value of $e_{cn}$ should account for the full angular dependence of the measured restitution.

\section{Conclusions}

This paper has addressed a long-standing difficulty in the modelling of non-spherical particle contact: how to control energy dissipation in a manner that remains consistent with the evolving contact dynamics. The central obstacle is that, unlike spherical contact, both the effective stiffness and the effective inertia depend on the contact configuration, so that the interaction cannot be reduced to a fixed one-dimensional oscillator. Classical damping formulations therefore fail not because they are poorly calibrated, but because they are structurally incompatible with the dynamics they are intended to represent.

Starting from this observation, the paper has developed a contact-centric reformulation in which the interaction is described directly at the contact point. This reveals two key features of non-spherical impact: a configuration-dependent projected normal mass $m_n^*$, which evolves during the collision as a breathing mass, and an intrinsic coupling between translational and rotational motion through the effective mass tensor $\mathbf{M}_{\text{eff}}$. These two effects are precisely what render conventional damping prescriptions inadequate.

To construct a consistent dissipative description, the paper then adopted the energy-phase transformation developed in earlier work. This transformation shows that any monotone contact potential can be mapped exactly onto a linear oscillator in transformed variables, and that precise control of energy loss requires the damping law to preserve this structure. The resulting damping form is therefore not empirical, but uniquely constrained by the requirement of velocity-independent restitution in the transformed system.

The main conceptual consequence is that the coefficient of restitution for non-spherical particle contact should be defined at the contact point. The appropriate quantity is the restitution of the contact-point normal velocity $e_{cn}$, rather than the total energy restitution $e_E$, which combines material dissipation with geometry-induced energy transfer. For non-spherical particles, the apparent variability of restitution with impact configuration is therefore not evidence of changing material behaviour, but a consequence of translational--rotational coupling during impact.

The numerical results support this interpretation. Across a wide range of aspect ratios, impact angles, and practically relevant stiffness levels, the proposed scheme controls $e_{cn}$ to within $1\%$ while reproducing the predicted variation of $e_E$ with geometry. The impulsive limit provides an exact closed-form account of the coupling effect, and the quality of the instantaneous approximation is shown to be governed by the single dimensionless parameter $\delta_{\max}/\rho$.

Taken together, these results establish a consistent framework for the control of energy dissipation in non-spherical particle contact. More fundamentally, they show that dissipation control and restitution definition cannot be separated: once the coupled contact dynamics are formulated correctly, the appropriate notion of restitution follows as a necessary consequence. 


\appendix

\section*{Appendix: Ellipsoid-wall contact geometry}
\label{sec:appendix}

For an ellipsoid with shape matrix $\mathbf{Q} = \mathbf{R}\,\text{diag}(a^{-2}, b^{-2}, c^{-2})\,\mathbf{R}^T$ at height $h$ above a wall at $z=0$, all quantities are computed from the Schur complement:
\begin{equation}
	Q_{\text{eff}} = Q_{33} - Q_{13}^2/Q_{11}, \qquad \rho = 1/\sqrt{Q_{\text{eff}}}.
\end{equation}
The cross-sectional area, cap volume, and contact centroid are:
\begin{equation}
	S_n = \frac{\pi}{\sqrt{Q_{11}Q_{22}}}\left(1 - \frac{h^2}{\rho^2}\right), \quad V_c = \frac{\pi}{\sqrt{Q_{11}Q_{22}}}\left(\frac{2\rho}{3} - h + \frac{h^3}{3\rho^2}\right), \quad x_c = \frac{Q_{13}\,h}{Q_{11}}.
\end{equation}
These satisfy $dV_c/dh = -S_n$. The lever arm is $\mathbf{r} = (x_c, 0, -h)^T$, from which $\mathbf{M}_{\text{eff}}^{-1}$ and $m_n^*$ follow via eqs.~(\ref{eq:M_eff}) and (\ref{eq:mn*}).

%

\bibliographystyle{elsarticle-num}
\bibliography{../../references}

\end{document}